## ARTICLE  OPEN

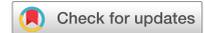

# Anomalous superconducting proximity effect of planar Pb–RhPb$_2$ heterojunctions in the clean limit

Rui-Feng Wang [1], Yan-Ling Xiong[1], Qun Zhu[1], Ming-Qiang Ren[2], Hang Yan[1], Can-Li Song [1,3✉], Xu-Cun Ma[1,3✉] and Qi-Kun Xue[1,2,3,4✉]

Interest in superconducting proximity effect has been revived by the exploitation of Andreev states and by the possible emergence of Majorana bound states at the interface. Spectroscopy of these states has been so far restricted to just a handful of superconductor-metal systems in the diffusion regime, whereas reports in otherwise clean superconductor-superconductor heterojunctions are scarce. Here, we realize molecular beam epitaxy growth of atomically sharp planar heterojunctions between Pb and a topological superconductor candidate RhPb$_2$ that allows us to spectroscopically image the proximity effect in the clean limit. The measured energy spectra of RhPb$_2$ vary with the spatial separation from proximal Pb, and exhibit unusual modifications in the pairing gap structure and size that extend over a distance far beyond the coherence length. This anomalously long-range proximity (LRP) effect breaks the rotational symmetry of Cooper pair potential in real space and largely deforms the Abrikosov vortex cores. Our work opens promising avenues for fundamental studies of the Andreev physics and extraordinary states in clean superconducting heterojunctions.

## INTRODUCTION

Interfacing quantum materials with superconducting condensates routinely leads to a leakage of Cooper pairs through the heterointerfaces *via* Andreev reflection[1–3], well-known as the proximity effect. This strategy has been emerging as an attractive route for the design of superconducting electronics and engineering of novel quantum states. If the proximitized materials are magnetic, topologically nontrivial or exhibit strong spin–orbit interaction[4–12], the long-sought Majorana bound states can emerge as non-Abelian anyons, potentially useful as the building blocks for fault-tolerant quantum computation[13]. In principle, the leakage of Cooper pairs induces subgap structures and profoundly modifies the local density of states (DOS) in the proximitized materials over a certain distance $d$ from the heterointerfaces[14–16], which has been widely explored with spatially resolved scanning tunneling microscopy/spectroscopy (STM/STS) techniques[17–25]. However, most of these studies are limited to vertically stacked superconducting heterojunctions[17–22], complicated by the interface strain control and discrete spatial variation. A continuous measure of superconducting subgap states and the proximity region with minimal strain necessitates a controllable synthesis of atomically sharp planar heterojunctions, which remains challenging in experiment[23–25].

In theory, the leaking distance of Cooper pairs in superconductor-metal heterojunctions is primarily governed by superconducting coherence length ($\xi$) and electron mean free path ($l$) in the normal metal[1,3]. Limited by the quality of the two constituent materials and the interface between them, previous heterojunctions used to explore the superconducting proximity effect were mostly in the diffusion regime (dirty limit) with $l < \xi$, where the leak of Cooper pairs takes place on a scale of the phase-coherence length in the diffusive metals[17–20,23,24]. This agrees markedly with a self-consistent solution of the Usadel equations[26,27]. On the other hand, LRP effects, which survive over scales largely exceeding the phase-coherence length of proximitized materials, have been unexpectedly evidenced in a large variety of systems and attracted increasing attention[25,28–34]. This is because they not only shed light on the superconducting pairing mechanism but also help develop high-performance superconducting devices. The underlying cause for the so-called LRP effects or giant proximity effects is phenomenally related to the properties of the proximitized materials that are either located in the ballistic regime (clean limit) with $l > \xi$[35] or characteristic of superconducting fluctuations[36], or both[33]. From this point of view, planar heterojunctions between two dissimilar superconductors are likewise desired for spatially resolved spectroscopic studies and microscopic understanding of the unpredicted LRP effects. To the best of our knowledge, only a few such experiments were previously reported by depositing small superconducting Pb islands on striped incommensurate Pb/Si(111) monolayer superconductors[37–39]. However, the proximity effect observed follows the conventional Usadel theory in the diffusion regime and was complicated by the geometric configuration of Pb islands[37]. A spatially resolved spectroscopy of the LRP effect has hitherto been lacking.

Two more general yet technically challenging issues are the spatial arrangement and core structure of the Abrikosov vortices within the proximity regions. Apparently, they are more relevant to hybrid heterojunctions with an LRP effect, in which the proximity regions are large enough to accommodate the Abrikosov vortex cores with a length scale of $\xi$. As the external magnetic field is applied normal to the hybrid heterointerface, the proximity vortex cores essentially mimic the superconducting ones but enlarge in size[10,40], whereas, thus far, neither theoretical nor experimental studies have been conducted to examine the impact of a magnetic field parallel to the heterointerface on the proximity vortices. Intuitively, the internal structure of vortex cores and even the spatial vortex arrangement might be radically modified by the spatial inhomogeneities of DOS and pair potential $\Delta$ within the proximity regions, but direct experimental evidence has been missing so far.

[1]State Key Laboratory of Low-Dimensional Quantum Physics, Department of Physics, Tsinghua University, Beijing 100084, China. [2]Southern University of Science and Technology, Shenzhen 518055, China. [3]Frontier Science Center for Quantum Information, Beijing 100084, China. [4]Beijing Academy of Quantum Information Sciences, Beijing 100193, China. ✉email: clsong07@mail.tsinghua.edu.cn; xucunma@mail.tsinghua.edu.cn; qkxue@mail.tsinghua.edu.cn

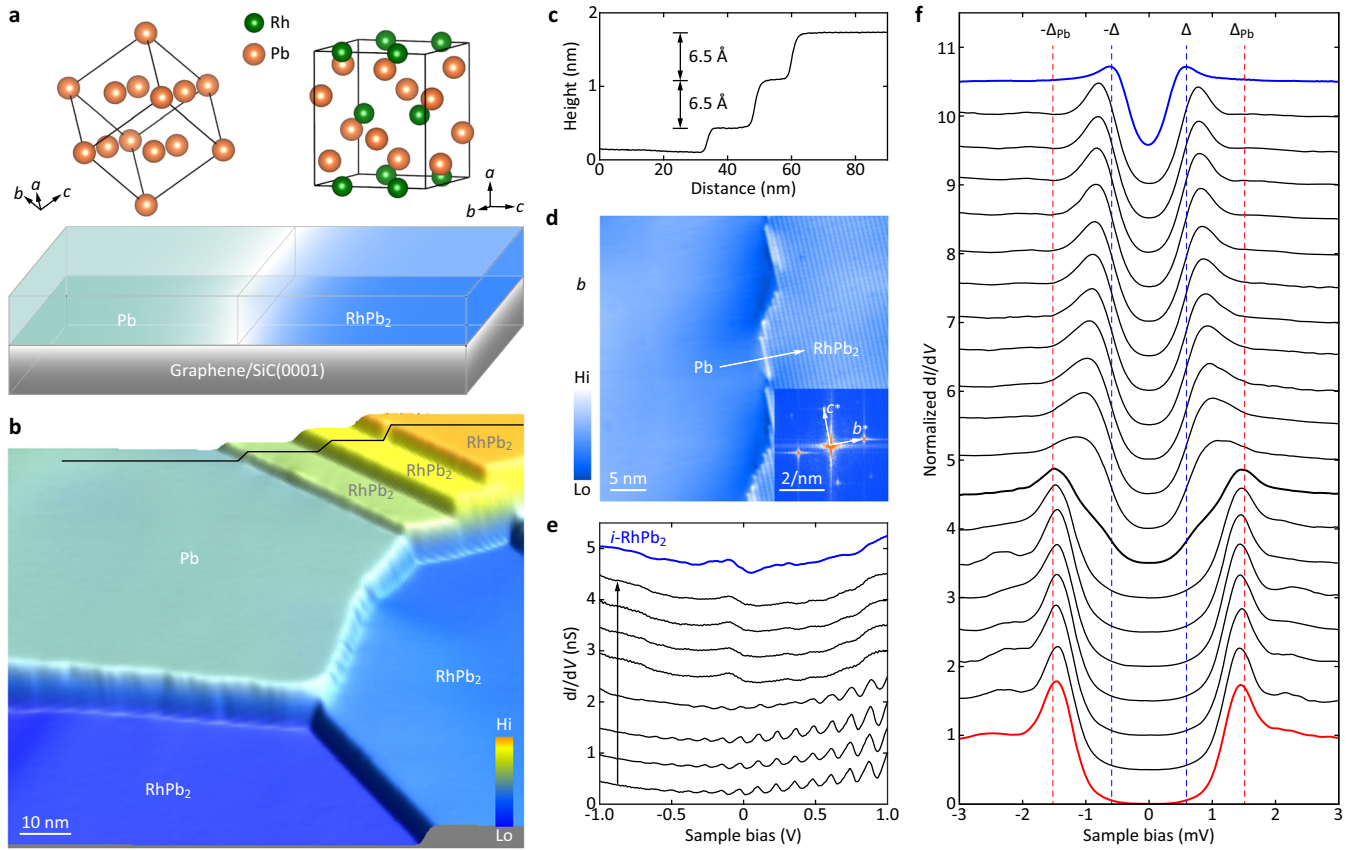

**Fig. 1 Planar heterojunctions between two dissimilar superconductors Pb and RhPb2. a** Schematic sketch of a planar Pb-RhPb$_2$ heterojunction on graphitized SiC(0001) substrate, with the respective crystal structure shown in the top panel. **b** Three-dimensional STM topographic image (100 nm × 100 nm, $V = 4.0$ V, $I = 5$ pA) of a representative Pb–RhPb$_2$ tricrystal heterojunction. As indicated, the Pb and RhPb$_2$ superconductor films are well connected and form atomically sharp heterointerfaces. **c** Topographic profile measured along the black trajectory across various terraces in (**b**). **d** Zoom-in STM topography (40 nm × 40 nm, $V = 12$ mV, $I = 1$ nA) showing the atomically sharp heterointerface between Pb and RhPb$_2$. Insert shows a FFT image of the $p$-RhPb$_2$ surface, from which the in-plane lattice parameters of $b = 6.8$ Å and $c = 5.9$ Å are estimated by measuring the 2D reciprocal vectors $b^*$ and $c^*$. **e** A series of d$I$/d$V$ spectra (setpoint: $V = 1.0$ V, $I = 500$ pA) taken along the arrowed line from Pb to RhPb$_2$ in (**d**). The topmost d$I$/d$V$ spectrum is acquired on $i$-RhPb$_2$ for comparison. **f** Low-energy-scale d$I$/d$V$ spectra (setpoint: $V = 4$ mV, $I = 500$ pA) acquired at equal separations (4.6 nm), illustrating a spatial evolution of the superconducting gaps across a Pb/RhPb$_2$ heterointerface. The bold curves are characteristic of $i$-RhPb$_2$ (blue), the Pb/RhPb$_2$ interface (black) and proximitized Pb (red), respectively. Vertical dashes mark the energy positions of the superconducting coherence peaks of intrinsic Pb (±$\Delta_{Pb}$, see the red ones) and $i$-RhPb$_2$ (±$\Delta$, blue ones) at 0.4 K. The spectra in **e** and **f** have been offset vertically on the d$I$/d$V$ intensity axis for clarity.

In order to answer the fundamental questions above, we managed to grow atomically sharp planar heterojunctions between two dissimilar superconductors, the conventional Pb and orthorhombic RhPb$_2$ single crystalline islands, on bilayer graphene/SiC(0001) substrates by the state-of-the-art molecular beam epitaxy (MBE) technique (see "Methods"). The geometry is schematically drawn in Fig. 1a. These heterojunctions enable us to probe continuously the energy-dependent DOS and local $\Delta$ via in situ STM/STS in proximitized RhPb$_2$ (hereinafter abbreviated as $p$-RhPb$_2$) islands with high spatial and energy resolution. Compared with the inherent RhPb$_2$ ($i$-RhPb$_2$) isolated from Pb, the $p$-RhPb$_2$ turns out to be anomalously modified in both superconducting gap structure and magnitude that extend over the entire visible RhPb$_2$ island explored, with a lateral dimension (up to hundreds of nanometers) far beyond the coherence length $\xi$ of RhPb$_2$. This exotic LRP effect, together with the unusual energy dependence of the decay length of subgap states away from the heterointerface, violates the conventional dirty-limit Usadel theory. As a magnetic field is applied perpendicular to the sample surface and therefore parallel to the Pb/RhPb$_2$ interface, we visualize Abrikosov vortex cores and zero-bias conductance peak (ZBCP) within vortices of RhPb$_2$, a hallmark of clean-limit superconductivity studied. Moreover, the ZBCP exhibits no visible splitting when moving away from the vortex center, hinting at a possible candidate of topological superconductor for RhPb$_2$. The Abrikosov vortex cores of $p$-RhPb$_2$ films are revealed to be deformed by a rotational symmetry breaking of the pairing magnitude $\Delta$ in real space. The present work offers a promising system for the fundamental studies of Andreev physics and extraordinary ground states in the clean superconducting heterojunctions.

## RESULTS

### Planar Pb–RhPb2 heterojunctions

Co-deposition of high-purity Pb and Rh on graphitized SiC(0001) substrates results in spontaneous epitaxial growth of single crystal-line Pb and Rh–Pb islands with heights of 15–30 nm. By combining X-ray diffraction (XRD) and in situ STM measurements, single-phase RhPb$_2$ binary films with a preferential orientation along (100) plane have been readily observed at lower substrate temperature $T_{sub} \leq 200$ °C, as detailed in Methods and Supplementary Fig. 1.

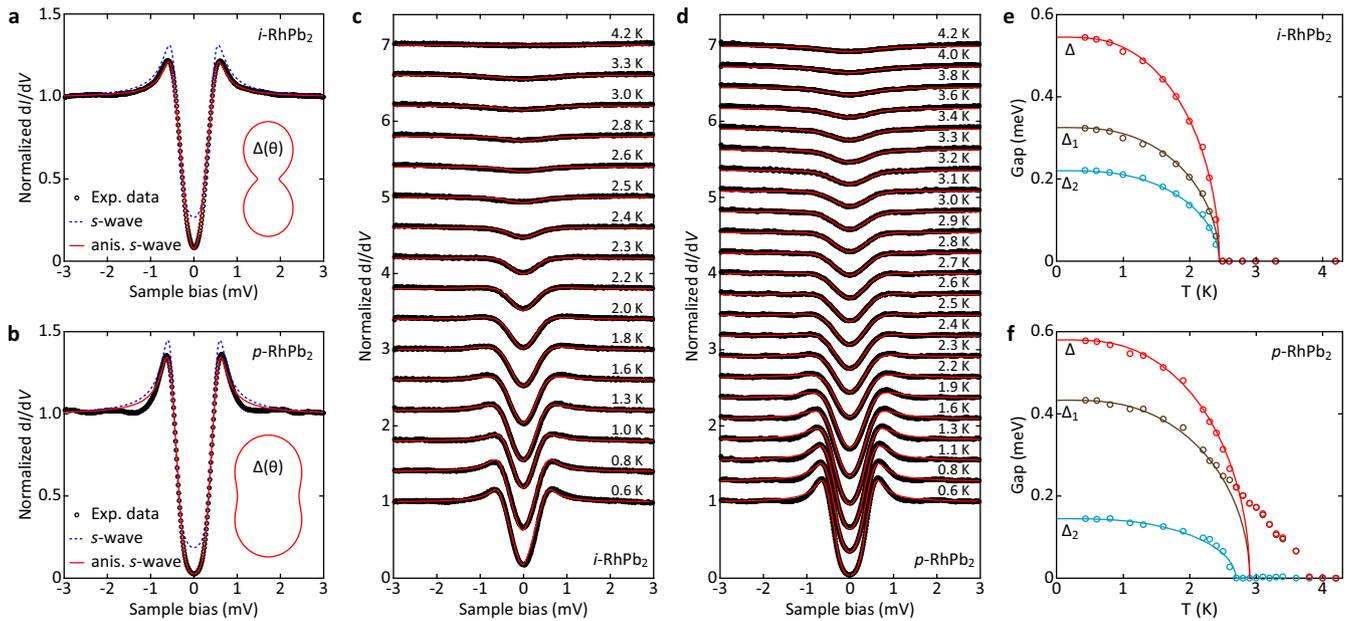

**Fig. 2 Temperature-dependent d$I$/d$V$ spectra of RhPb2 without/with proximal superconductor Pb. a, b** Normalized d$I$/d$V$ spectra (setpoint: $V = 3$ mV, $I = 500$ pA) on $i$- and $p$-RhPb2 at 0.4 K, respectively. The colored lines show the best fits of the experimental data (empty circles) to the Dynes model with isotropic (blue dashed lines) and anisotropic (red solid lines) $s$-wave gap functions. Insets show the polar plot of the superconducting gap $\Delta$ as a function of $\theta$. The normalization was performed by dividing the raw tunneling d$I$/d$V$ spectra by their backgrounds, extracted from a linear fit to the conductance beyond the superconducting gaps. **c, d** Dependence of d$I$/d$V$ spectra on temperature as indicated on both $i$- and $p$-RhPb2, respectively. The red lines describe the best fits to anisotropic $s$-wave gap functions. The spectra have been offset vertically on the d$I$/d$V$ intensity axis for clarity. **e, f** Temperature-dependent superconducting gap parameters of $\Delta$ (red), the isotropic component $\Delta_1$ (brown) and anisotropic component $\Delta_2$ (cyan) on $i$- and $p$-RhPb2 films, respectively. Solid lines represent the fitting by the BCS gap function.

It is further found that the RhPb2 phase adopts an orthorhombic crystal structure with lattice constants of $a = 6.5$ Å, $b = 6.8$ Å and $c = 5.9$ Å (Fig. 1b–d) and differs from the previously synthesized RhPb2 alloy superconductor with a tetragonal CuAl2-type (space group: I4/mcm) structure[41]. More intriguingly, the orthorhombic RhPb2 islands enable themselves to be seamlessly connected with the (111)-oriented Pb islands (Fig. 1a, b, d), thereby forming well-coupled, atomically sharp planar heterojunctions, despite the lack of a fixed in-plane epitaxial alignment between the two dissimilar superconductors of Pb and RhPb2.

Figure 1b shows a three-dimensional (3D) topography of the typical Pb/RhPb2 heterojunctions with regular step-terrace structures. The line profile across various terraces justifies the sharpness of the heterointerface and a monoatomic step height of 6.5 Å for RhPb2 (Fig. 1c), in good agreement with the lattice parameter $a$ measured from XRD (Supplementary Fig. 1). The atomically sharp Pb/RhPb2 heterointerface has been further confirmed by a magnified STM topography in Fig. 1d. Although the proximitized Pb surface is somewhat prone to get corrugated, the atomically resolved RhPb2 is commonly characteristic of a rectangular 2D lattice with $b = 6.8$ Å and $c = 5.9$ Å, determined by a fast Fourier transform (FFT) algorithm (see the inserted FFT image in Fig. 1d). Notably, we can also obtain $i$-RhPb2 islands separated from proximal Pb with the same lattice constants by reducing the atomic flux ratio between Pb and Rh (<3.4) during the co-deposition growth. This provides us an opportunity for comparative spectroscopic studies of $i$-RhPb2 and $p$-RhPb2.

Plotted in Fig. 1e are a series of tunneling differential conductance (d$I$/d$V$) spectra across the Pb–RhPb2 heterointerface (black curves) as well as one representative spectrum measured directly on $i$-RhPb2 (blue curve) for comparison. The spectra are quite uniform on respective sides, but display distinct features and a well-defined boundary between Pb and RhPb2. Specifically, the orthorhombic RhPb2 films, regardless of their proximity to Pb, are characteristic of metal-like DOS with a visible hump at around −0.1 eV, while the Pb films exhibit discrete quantum well states by the quantization of their electronic states[42]. According to the energy separation between adjacent states around the Fermi level ($E_F$), the film thickness is estimated to be ~ 29 nm (100 monolayers). Figure 1f illustrates low-energy d$I$/d$V$ spectra across one Pb/RhPb2 heterointerface in Fig. 1b, from which a continuous variation of the superconducting energy gaps from the proximitized Pb to RhPb2 films (from bottom to top) has been readily resolved in real space. At first glance, the most prominent feature develops near the very interface of <1 nm (see also Supplementary Fig. 2) where the superconducting gaps are complicated by a mixed pairing sign of Pb and RhPb2 (see the bold black curve in Fig. 1f). These observations corroborate spectroscopically the well-coupled and atomically sharp heterojunctions between the Pb and RhPb2 films. Moreover, we reveal that the superconducting energy gaps of RhPb2 films change little with the thickness (> 10 nm), but only rely on the distance away from the proximal Pb islands. This excludes the possibility of any Pb beneath the epitaxial RhPb2 films.

## Proximity-modified superconducting gaps and the temperature dependence

A close examination reveals enhanced and suppressed superconducting energy gaps in the proximitized RhPb2 (Figs. 1f, 2a, b and Supplementary Fig. 3) and Pb films (Supplementary Fig. 4) with respect to the bare ones, respectively, hallmarks of the proximity effect. Unexpectedly, all $p$-RhPb2 films are observed to be anomalously modified in gap structure and magnitude that extend over the entire visible RhPb2 islands enclosed, with a lateral dimension far exceeding the coherence length $\xi$ of ~ 28 nm in RhPb2. Note that $\xi$ is deduced by exploring the magnetic field $B$ dependence of superconducting gaps in $i$-RhPb2, as explained in Supplementary Fig. 5. Such a finding runs counter to the conventional dirty-limit Usadel theory and instead points to a LRP effect. Beyond the qualitative distinction of $\Delta$, the superconducting gap structure reveals striking contrasts between the $i$-RhPb$_2$ and $p$-RhPb$_2$, by fits of the normalized d$I$/d$V$ spectra to a well-known Dynes mode with isotropic and anisotropic $s$-wave gaps[43]. Given the orthorhombic crystal structure of RhPb$_2$, a twofold-symmetric anisotropic $s$-wave function $\Delta(\theta) = \Delta_1 + \Delta_2\cos2\theta$ instead of a fourfold-symmetric one has been assumed ($\theta$ is the in-plane angle), although they actually give the same gap shape. Whereas, in $i$-RhPb$_2$ (Fig. 2a), the experimental d$I$/d$V$ spectrum matches nicely with a highly anisotropic $s$-wave superconducting gap of $\Delta(\theta) = 0.55(0.60 + 0.40\cos2\theta)$, the gap anisotropy is obviously reduced in $p$-RhPb$_2$, with $\Delta(\theta) = 0.59(0.75 + 0.25\cos2\theta)$ at a distance of about 76 nm from the proximal Pb (Fig. 2b). Note that the gap maximum $\Delta \sim 0.55$ meV in $i$-RhPb$_2$ is smaller than the isotropic energy gap $\Delta_{Pb} \sim 1.37$ meV in Pb. As a result, considerable subgap states develop in the energy interval $E \in [\Delta_1 - \Delta_2, \Delta_{Pb}]$ (Supplementary Fig. 4), which, together with the suppressed coherence peaks, renders it nontrivial to quantitatively fit the experimental tunneling d$I$/d$V$ spectrum in proximitized Pb. Rather than being limited to the pair magnitude, as routinely investigated before in ordinary superconducting heterojunctions, our results stress that the gap structure is subject to change in heterojunctions comprising dissimilar superconductors.

Having established the proximity-induced modifications in the pairing gap structure and magnitude, we now turn to the temperature dependence in both $i$- and $p$-RhPb$_2$. Fits of the Dynes mode for all temperature-dependent superconducting gaps in Fig. 2c, d (red curves) allow us to extract readily the corresponding gap values of $\Delta_1$, $\Delta_2$ and $\Delta = \Delta_1 + \Delta_2$, which are described in Fig. 2e, f, respectively. Evidently, all superconducting gaps of $\Delta_1$, $\Delta_2$ and $\Delta$ nicely follow the conventional BCS gap function and vanish at the same transition temperature $T_c = 2.45$ K (Fig. 2e)[44], slightly smaller than $T_c \sim 2.66$ K reported in the tetragonal RhPb$_2$ phase[45]. The reduced gap ratio is estimated to be $2\Delta/k_BT_c = 5.2$ that exceeds the canonical BCS value of 3.53 ($k_B$ is the Boltzmann constant), indicative of a strongly coupled superconductivity for the orthorhombic RhPb$_2$ phase. However, the temperature dependence of the gap values in $p$-RhPb$_2$ proves more complicated than ever anticipated (Fig. 2f). Firstly, the anisotropic component $\Delta_2$ reduces to zero at 2.7 K (only slightly larger than $T_c$ of $i$-RhPb$_2$), but the isotropic component $\Delta_2$ and thus $\Delta$ persist up to a higher temperature of 3.8 K. Obviously, the residual minigap above $T_c = 2.90$ K in $p$-RhPb$_2$ is related to its proximity to Pb and reminiscent of that identified in finite normal metal-superconductor heterojunctions[16,19,27,40,46]. In addition, the reduced anisotropy of $\Delta$ does not derive trivially from a simple increase of its isotropic component $\Delta_1$ due to the proximal Pb, but also from an unexpected decrease of the anisotropic component $\Delta_2$ (Fig. 2e, f).

## Spatially resolved proximity effect

To fully understand the LRP effects, we have measured the site dependence of the superconducting gaps in one representative Pb/RhPb$_2$ heterojunction (Fig. 3a), with our primary focus on the $p$-RhPb$_2$ surface. Two spatially resolved spectroscopic d$I$/d$V$ maps at $E = 0$ meV and 0.3 meV, shown in Fig. 3b, d, respectively, exhibit monotonic and quasi-1D DOS modulation normal to the hetero-interface, with no sign of geometry-controlled proximity effect as observed in the heterogeneous superconductor thin films of Pb[37]. This behavior has been similarly found in another tricrystal junction, where two orientation-distinguishing RhPb$_2$ islands are closely connected with Pb (Supplementary Fig. 2). Moreover, the DOS reaches no saturation at a distance of >100 nm away from the interface, again in contrast to the expectation by the dirty-limit Usadel theory, where the proximity DOS vanishes over a length scale of $\xi^{17-20,23,24}$.

Taking advantage of the geometry-free proximity effect, we can accurately characterize the spatial dependence of the proximity-induced DOS on the distance $d$ away from the heterointerface. These color-coded tunneling d$I$/d$V$ spectra acquired at 0.4 K (Fig. 3d, e) and 4.2 K (Fig. 3f, g) emphasize the sharp boundary (marked by the white dashed lines at $d = 0$) between the Pb and RhPb$_2$ islands as well as the enhanced superconducting gaps that extend to distances anomalously longer than $\xi$ below $T_c$ of RhPb$_2$. For simplicity, we measure the energy separation between the two coherence peaks ($E_{peak}$: slightly larger than $2\Delta$ due to the thermal smearing) and the energy-dependent DOS within the energy gap $\Delta$ to quantify the spatial variations of the superconducting gaps (Fig. 3h, i). All of them decay exponentially with $d$ (see the solid red and black lines). The decay lengths (or saying the diffusion length $\delta_E$ of Andreev quasiparticles) are estimated to be $\delta_E \sim 27$ nm for $E_{peak}$ at 0.4 K (Fig. 3h) and $\sim 20 \pm 3$ nm for all DOS($E$) at 4.2 K (bottom panel in Fig. 3i), close to slightly smaller than the coherence length $\xi \sim 28$ nm in RhPb$_2$, respectively. However, the $\delta_E$ of DOS at 0.4 K turns out to be extremely larger (>200 nm for lower $E$) than $\xi$ and presents a divergent increase as the energy $E$ approaches to zero (Fig. 3j). Note that the revealed decay of $\delta_E$ with energy, namely $\delta_E \propto 1/E^{1.46 \pm 0.07}$, is in striking contrast to the behavior of normal metal-superconductor heterojunctions, where the diffusion length $\delta_E$ is given by $\delta_E = \sqrt{\hbar D/E}$ with $D$ marking the diffusion constant[2,38,40]. A tiny discontinuity in $E_{peak}$ and DOS at 0.4 K suggests a highly transparent yet nonperfect interface studied (Fig. 3h and top panel of Fig. 3i), while a DOS peak anomaly occurs justly on the interface at 4.2 K (bottom panel of Fig. 3i) and its cause remains unknown.

## Proximity vortices in $p$-RhPb$_2$

Single or multiple quantized Abrikosov vortices have been observed, by mapping the spatial distribution of zero-bias conductance (ZBC) under moderate magnetic field $B$, to penetrate the proximity regions of $p$-RhPb$_2$, as outlined by the white square in Fig. 4a. In Fig. 4b, a single Abrikosov vortex behaves itself as a comet-like petal (orange region, see also the purple contour lines for guide) with enhanced ZBC due to the suppression of superconductivity there, with the principal symmetry axis normal to the interface. This contrasts markedly with nearly circular vortex cores in $i$-RhPb$_2$ (Fig. 4c and Supplementary Fig. 6). The distinctly deformed Abrikosov vortex cores have been compellingly corroborated in the proximity regions of all $p$-RhPb$_2$ films explored. In heterojunctions that are large sufficient to accommodate multiple vortices in $p$-RhPb$_2$ below $B_{c2}$ (Fig. 4d and Supplementary Fig. 7), the Abrikosov vortex cores are robustly elongated along the direction normal to the Pb/RhPb$_2$ hetero-interface, and the vortex-induced bound states at vortices present prominent orientation dependence (Fig. 4e).

One more important finding stands out by visualizing the vortex-induced bound states in RhPb$_2$. In Fig. 4f, we show emergence of a pronounced zero-bias conductance peak (ZBCP) that progressively weakens with increasing field $B$ at the center of vortex core in $i$-RhPb$_2$. This means a longer $l$ than $\xi$ (clean limit) in which the constructive interference of repeated Andreev scatterings leads to the salient ZBCP within vortices, except that the Abrikosov vortices are pinned to the heterointerfaces (Supplementary Fig. 7). Likewise, the clean-limit superconductivity holds true for $p$-RhPb$_2$ (Fig. 4g). We here ascribe it to the rare realization of atomically sharp interfaces between the epitaxial RhPb$_2$ films and chemically insert graphene substrates, which have significantly minimized the interfacial scattering. This result distinguishes the present RhPb$_2$/Pb system from most of the dirty-limit superconducting heterojunctions previously studied[17–20,23,24,37–39]. More strikingly, the ZBCP is found to be invariably fixed to $E_F$, albeit suppressed amplitude, when moving away from the center of magnetic vortices until the superconducting gap recovers at a radial distance $\xi$ (Fig. 4g and

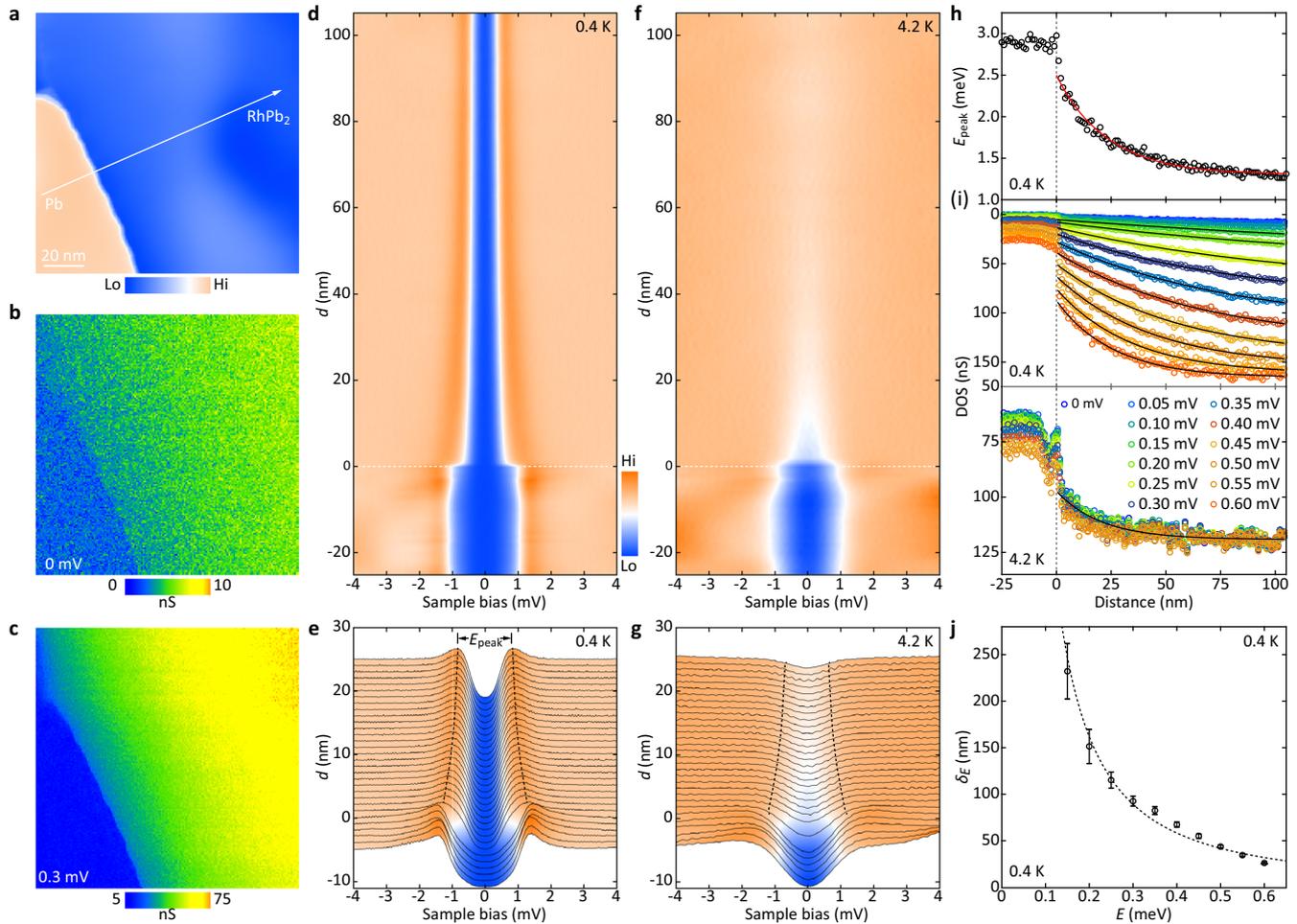

**Fig. 3 Spatially resolved proximity effect. a** STM topography (130 nm × 130 nm, $V = 90$ mV, $I = 10$ pA) of another Pb-RhPb$_2$ planar heterojunction, with the Pb island slightly higher (7 Å) than RhPb$_2$. **b, c** Differential conductance d$I$/d$V$ maps recorded simultaneously with image (**a**) at energies of zero and 0.3 meV, respectively. **d** Color-coded and distance $d$-dependent d$I$/d$V$ spectra (setpoint: $V = 4$ mV, $I = 500$ pA) measured at 0.4 K along the arrowed line (normal to the Pb/RhPb$_2$ heterointerface) in (**a**). The spectra are taken at equal separations (1 nm), while the white dashes denote the interface with $d = 0$. **e** 3D view of the d$I$/d$V$ spectra near the very heterointerface in (**d**). The black dashed lines mark the spatial variation of coherence peaks. **f, g** The same data as in (**d, e**), but for 4.2 K. **h** Spatial evolution of $E_{peak}$ by measuring the energy separation between the two coherence peaks in (**d**). Red line shows the best fit of $E_{peak}$ to an exponential law, yielding a decay length of ~ 27 nm. **i** Spatial and energy variations of the low-lying electronic DOS within the superconducting gaps of $p$-RhPb$_2$ at 0.4 K (top panel) and 4.2 K (bottom panel), with the black curves showing the exponential fits. **j** Energy-dependent decay length $\delta_E$ for the low-lying DOS in $p$-RhPb$_2$, exhibiting a divergent increase towards zero energy at $T = 0.4$ K. The error bars show the standard deviations of $\delta_E$.

Supplementary Figs. 6e, 7f). It should be emphasized that the spatial non-splitting of ZBCP is unusual and inherent to RhPb$_2$, no matter whether being in close proximity to the superconducting Pb films. Note that a minigap occurs above $B_{c2}$ as well (Fig. 4f and Supplementary Fig. 7d) that probably relates to a geometrical confinement effect in the nanosized RhPb$_2$ islands[46].

## DISCUSSION

Our observation of the spatially non-split ZBCP in an alloy superconductor RhPb$_2$ sharply differs from the usual Caroli-de Gennes–Matricon (CdGM) vortex bound states[47], which have been generally revealed to split at sites away from the vortex center as predicated[48–50]. Alternatively, if a Majorana zero mode (MZMs) is involved and intermixed with the CdGM states, the ZBCP splitting will be significantly delayed[10,51,52] or even never occur within the Abrikosov vortices[53–56]. Under this context, our intriguing finding hints at a possible candidate of innate topological superconductor for RhPb$_2$ that harbors MZMs within vortices. It sounds reasonable since both Rh and Pb are heavy elements granting the strong spin-orbital coupling. In theory, a tetragonal allotrope β-RhPb$_2$ (space group: I4/mmm) has been recently predicted to be a potential topological superconductor just like β-PdBi$_2$[53,57–60]. Therefore, it is highly interesting to further investigate the topological nature of the orthorhombic RhPb$_2$ and the non-split ZBCP observed here.

Anyhow, a critical question naturally arises as to whether the LRP effect, namely the anomalous long decay length $\delta_E \gg \xi$ of low-lying DOS within the superconducting gaps at 0.4 K (Fig. 3i) and the modified superconducting gap structure in the entire visible $p$-RhPb$_2$ islands (Fig. 2), correlates with the possible topological surface states for ballistic transport in $p$-RhPb$_2$[22,25,34]. Given that the LRP effect is absent at 4.2 K, a temperature between the $T_c$ values of Pb and RhPb$_2$ (Fig. 3f and bottom panel of Fig. 3i), this scenario may not hold true or not act as the sole cause of the observed LRP effect in Pb/RhPb$_2$. Alternatively, it implies the essential role of the preexisting Cooper pairs in stabilizing the LRP effect at 0.4 K (Fig. 3d, h and top panel of Fig. 3i). Provided that the LRP effect is missing in the dirty-limit heterojunctions of Pb islands and superconducting Pb monolayer films[38], our findings unravel that the LRP effects arise from the cooperative effects of preexisting Cooper pairs and ballistic transport in $p$-RhPb$_2$. Although the microscopic mechanism of LRP effect has not been well understood, we believe that the Cooper pairs tunneling across the atomically sharp Pb/RhPb$_2$ interface, multiple Andreev reflections and ballistic propagation of the carriers in finite-sized RhPb$_2$ are highly relevant. This has been supported by a careful inspection of the $p$-RhPb$_2$ and proximitized Pb films (Fig. 2 and Supplementary Figs. 3, 4), which present several dip-hump-like features outside the superconducting gaps, probably associated with the geometry-induced Tomash oscillation[61,62].

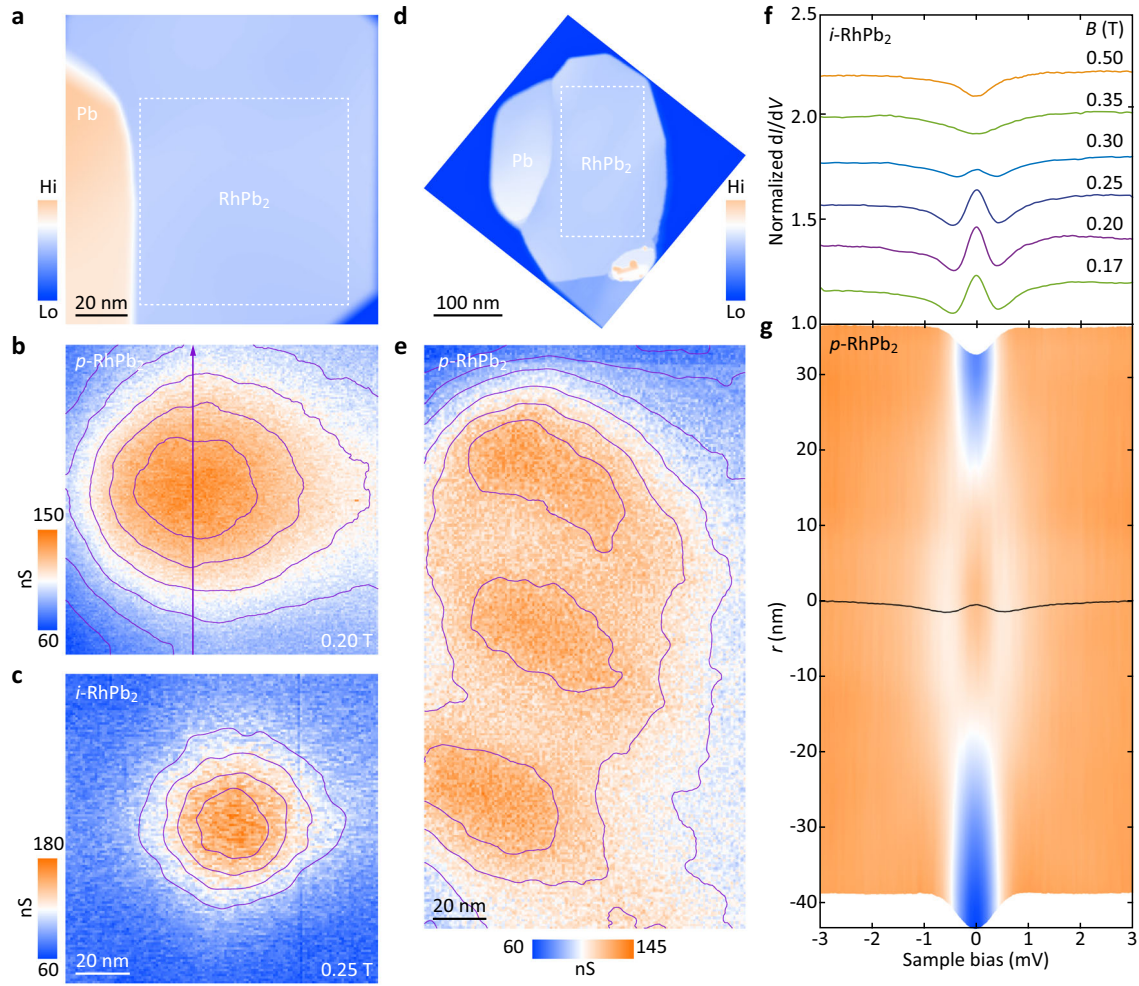

**Fig. 4 Proximity vortices. a** STM topography (120 nm × 120 nm, $V = 4$ V, $I = 5$ pA) of one Pb/RhPb$_2$ heterojunction where the proximity effect on magnetic vortices is studied. **b** ZBC map (80 nm × 80 nm, $V = 3$ mV, $I = 500$ pA) acquired in the square-outlined proximity region of (**a**), showing a petal-like Abrikosov vortex core at $B = 0.20$ T. The purple contour lines are guides for eyes throughout. **c** ZBC map (115 nm × 115 nm, $V = 3$ mV, $I = 500$ pA) showing a nearly circular vortex core in $i$-RhPb$_2$ ($B = 0.25$ T). **d** STM topographic image (300 nm × 300 nm, $V = 4$ V, $I = 5$ pA) of a larger Pb-RhPb$_2$ heterojunction. **e** ZBC map (120 nm × 220 nm, $V = 3$ mV, $I = 500$ pA) presenting three Abrikosov vortices at $B = 0.20$ T, acquired in the rectangle-outlined proximity region of (**d**). **f** d$I$/d$V$ spectra taken at the Abrikosov vortex center of $i$-RhPb$_2$ for various $B$. **g** Color-coded d$I$/d$V$ spectra measured with varying radial distance $r$ from the vortex center ($r = 0$, the black line) along the arrowed line in (**b**). A prominent ZBCP exists at vortices and shows no splitting away from the vortex center.

Finally, we comment on the deformed vortices in $p$-RhPb$_2$. Given the fact that the vortex core size $\xi$ is cooperatively determined by the Fermi velocity $v_F$ and superconducting energy gap $\Delta$, the gap anisotropy (Fig. 2a) suggests an anisotropic $v_F$ (compatible with the orthorhombic crystal structure) that compensates for the anisotropy of $\Delta$ to produce the nearly circular vortex cores in $i$-RhPb$_2$ (Fig. 4c and Supplementary Fig. 6). Therefore, the appearance of petal-like or strongly elongated vortices along the direction normal to the Pb/RhPb$_2$ interface arises from a geometry effect in the proximity regions of $p$-RhPb$_2$ (Fig. 4b, e), rather than any anisotropy of either $\Delta$ or $v_F$[63,64]. Distinct from a pure geometry confinement effect[65], however, the rotational symmetry breaking of the pair potential $\Delta$ in real space plays a decisive role in deforming the Abrikosov vortex cores of $p$-RhPb$_2$. Specifically, the strong LRP effect induces a 1D spatial variation of $\Delta$ that monotonically reduces with $d$ (>0, Fig. 3h). Consequently, the pair potential $\Delta$ and thus $\xi$ become site-dependent, thereby deforming the vortex cores with a ZBC tail away from the heterointerface to minimize the system energy (Fig. 3b, e). These anomalous experimental observations provide insights into the superconducting proximity effect and a promising strategy to further explore the Andreev and Majorana physics based on well-coupled, clean-limit heterojunctions between two dissimilar superconductors.

## METHODS

### Sample growth

All samples were grown by co-depositing high-purity Rh (99.95%) and Pb (99.999%) metal sources on graphitized SiC substrates via

MBE with a base pressure of $2 \times 10^{-10}$ Torr. Prior to the MBE growth, the double-layer graphene was obtained by heating the nitrogen-doped SiC(0001) wafers to about 1350 °C for 10 min using the well-established recipe[66], and fixed at 150–200 °C during the MBE growth to access high-quality, single-phase $i$-RhPb$_2$ and well-connected Pb/RhPb$_2$ heterojunctions. The Rh and Pb sources were well-outgassed and evaporated from a single-pocket electron beam evaporator (SPECS) and standard Knudsen diffusion cell, respectively. A fine control of the atomic flux ratio between Pb and Rh sources via a quartz crystal microbalance (QCM, Inficon SQM160H) can lead to the MBE preparation of many individually isolated $i$-RhPb$_2$ islands (Pb/Rh < 3.4) without proximal Pb and Pb/RhPb$_2$ hybrid heterojunctions (Pb/Rh ≥ 3.4).

**In situ STM measurements**

Our STM experiments were carried out in a commercial Unisoku USM 1300 $^3$He system, integrated with an MBE chamber for in situ sample preparation. Unless specified otherwise, all STM and STS measurements were conducted at 0.4 K with a polycrystalline PtIr tip, which was cleaned via $e$-beam bombardment in MBE and appropriately calibrated on MBE-grown Ag/Si(111) films. All STM topographic images were acquired in a constant current mode with the bias $V$ applied to the samples, while the tunneling d$I$/d$V$ spectroscopy and conductance map were measured using a standard lock-in technique with a small bias modulation at 931 Hz.

## ACKNOWLEDGEMENTS

This work was supported by the National Key Research and Development Program of China (Grant Nos. 2018YFA0305603 and 2017YFA0304600), the National Natural Science Foundation of China (Grant Nos. 51788104 and 11634007).

## AUTHOR CONTRIBUTIONS

C.L.S. X.C.M. and Q.K.X. designed the project. R.F.W. and Y.L.X. with the aid of Q.Z. M.Q.R. and H.Y. carried out MBE growth and STM/STS measurements. R.F.W. and C.L.S. analyzed the experimental data. R.F.W. C.L.S. and X.C.M. interpreted the results and co-wrote the paper with input from other authors.

## COMPETING INTERESTS

The authors declare no competing interests.

## ADDITIONAL INFORMATION

**Supplementary information** The online version contains supplementary material available at https://doi.org/10.1038/s41535-022-00529-4.

**Correspondence** and requests for materials should be addressed to Can-Li Song, Xu-Cun Ma or Qi-Kun Xue.

**Reprints and permission information** is available at http://www.nature.com/reprints

**Publisher's note** Springer Nature remains neutral with regard to jurisdictional claims in published maps and institutional affiliations.